# Assessing the Dimensions of Relationship Quality in B2C E-Banking Services: An Empirical Comparative Study

Ali Ahmad Alawneh

**Department of Management Information Systems, Philadelphia University**
**Amman, Jerash-Amman Road, 19392, Jordan**

**Abstract**
In B2C Online services contexts where relationships between customers and service providers matter due to the lack of face-to-face interactions, uncertainty, intangibility, hyper competition, increasing risk of fraud and lack of trust. Relationship quality (RQ) is replacing service quality as a key source of superior performance and competitive advantage. Accordingly, this study investigates the dimensions of relationship quality from e-Banking services and explores the differences among customers of Jordanian and foreign banks. Many of the foreign banks in Jordan are equipped with large financial capital, having high banking and financial experiences and are offering many modern hi-tech banking services. This situation makes Jordanian banks in front of unprecedented competition and so they have invested heavily to develop their e-banking services, in order to strengthening their relationships with existing customers, enhancing their trust, letting them satisfied, ensuring their commitment and attracting of new ones. This paper aims to identify the key dimensions that shape the relationship quality among Jordanian and foreign banks and their customers who are utilizing the e-Banking services. Based on an extensive review of relevant literature, we have formulated nine hypotheses and identified three factors (satisfaction, trust, and commitment) that may affect the competitiveness and success of Jordanian and foreign banks. Survey data from 350 customers from four Jordanian banks and four foreign banks in Amman city the capital of Jordan were collected and used to test the proposed hypotheses. Based on the structural equation modeling and T-test analyses, our empirical analysis demonstrates several key findings. These findings indicate the importance of investigation the dimensions of relationship quality so as to provide feedback in a set of recommendations that will form a basis to establish, develop and maintain successful and strong relationships which in turn will results in increased business performance of the banks.

## 1. Introduction

Rapid advances in information technology have dramatically changed how services are conceived and delivered (Massey et al. 2007). Based on technological advances, researchers and managers are acknowledging that the source of competitive advantage is closely related to the quality of long-term relationships between partners (Palmer 2002). Given that the product/service offered by companies in a given market may be essentially the same, differentiation is exerted through the capacity of developing long-term relationships with clients that resist changes in the competitive environment (Palmatier et al. 2006; Rauyruen and Miller 2007).

Information and Communication Technology (ICT) has transformed radically the way individuals, organizations, and governments used to work. In today's Information societies, Internet has become an essential channel that is used for disseminations of information, products, and services. People would like to use Internet as a transaction tool in different areas, such as, learning, shopping, marketing, travel, banking, trading, etc. Banks also have realized the importance of Internet and undertake critical transformation to use Internet to deliver online banking services so that customers access them whenever and wherever they want (Alawneh and Hattab, 2009).

Relationship marketing is mainly designed to increase customer coherence. The relationship between buyers and sellers is the key to successful business operations and is crucial to maintaining a competitive advantage. The development of global logistics systems increases the profit of traditional banks as Internet technologies are combined with information applications. In the virtual context of the Internet, customer loyalty is strongly related to his/her trust and satisfaction with an Internet bank (Palmatier et al. 2006).

In the service industry, and in particular in service industries that require more personal interaction and customized service, the interaction between customers and salespeople determines the ability to make deals. Trust means that behavior is predictable and reduces the risk of uncertainty. Mutual familiarity can break down psychological barriers. Consequently, trust serves as a barometer of interpersonal relationship quality. According to social exchange theory, the benefit is the award (or profit) gained from the interpersonal relationships minus the cost of the commodity. This benefit determines the satisfaction obtained from the relationship (Sun 2010).

Relationship quality is focused more on long-term customer relationships rather than on short-term transactions. Previous researchers have shown that a high quality relationship can earn life-long customer commitment and is valuable for a business. Such a relationship reduces uncertainty for customers and helps maintain a good business relationship. Crosby et al. (1990) suggested that relationship quality comprises two parts, trust and satisfaction, which are each considered an "emotional state that occurs in response to an evaluation of these interaction experiences".

When face-to-face meetings are difficult to arrange, the Internet technologies and applications (such as e-Banking) provide a low-cost and widely accessible medium that can be used to enhance the effectiveness of the bank to better understand the customer, and to promote its services effectively such that the business relationship improves. The effectiveness of various interactions can be enhanced by overcoming the barriers that arise due to the geographical distance in domestic or regional banking activities, and by taking advantage of the advances in the Internet technologies (such as e-Banking, home banking, phone banking, SMS banking, WAP banking, e-mailing, instant messaging, voice-over-IP, and e-channels).

Internet applications provide customers with timely service. Furthermore, banks can use relationship marketing to improve customer loyalty. High relationship quality indicates that online users trust Internet banks, are satisfied with the banking services, and perceive Internet banking to be secure. Additionally, a cooperative relationship can be created by establishing good interactions between banks and customers. The best ways to retain customers are satisfying customer needs and cultivating customer trust and loyalty. Numerous researchers have attempted to identify the various influences on customer retention. The key to business survival and success is building and maintaining strong customer relationships. Crosby et al. (1990) noted that services are intangible and complicated, and that customers have little knowledge of their nature. Message delivery can take a long time owing to high customer uncertainty regarding services. In order to build and retain a good sales relationship with customers, businesses must establish good quality relationships. A good relationship can satisfy customers, cultivate trust, and consequently reduce service uncertainty.

This study is interested in the relationship between the Jordanian and foreign banks that are providing e-banking services and their customers. In the context of e-Banking, uncertainty, urgent responsiveness, and competitiveness are very common. Thus the quality of the relationship with customers is critical for the success and distinguishes of the banks that are providing e-Banking services.

Both the banks that are providing e-Banking services and the customers can benefit from a high quality relationship. For the banks, maintaining a high quality relationship with its customers is critical for relationship marketing which aims at achieving cost-saving, enhancing relationships, attracting new customers, maintaining loyalty and providing a reliable source of future revenues and profits. For customers, a high quality relationship with the banks is also beneficial in that it helps customers make a long-term commitment which reduces the risk of future interactions and benefit in fulfillment of their expectations and demands. Therefore, both the banks that are providing e-Banking services and their customers are motivated to maintain a high quality relationship with each other.

The value of this research lies in the synthesis of the prior research on satisfaction, trust and commitment from a relationship quality perspective. This synthesis has the following advantages. First, it enriches our understanding of satisfaction, trust and commitment. From prior relationship quality research, we know that these three dimensions are rooted in marketing and customer relationships literature. The synthesis can further specify how relationship quality is reflected at different magnitudes by satisfaction, trust and commitment. Second, the synthesis of these three dimensions in relationship quality helps us to connect the capabilities of e-Banking services to business benefit factors more explicitly and effectively, given that that relationship quality has been closely tied with business benefit factors in marketing literature. Moreover, this research explores the role of e-Banking services in strengthening the relationships between banks and their customers in terms of enhancing trust, satisfaction and commitment. Finally, this research aims at investigating the differences among Jordanian and foreign banks regarding the three dimensions of relationship quality.

The structure of the paper proceeds as follows: Section two gives an overview of e-banking services in Jordan; Section three is a review of the key dimensions of relationship quality as found in the marketing literature; in Section four, an overview of previous and related works which provide the foundation for the current study is given; Section five provides the hypothesized relationships between relationship quality and its dimensions. This is followed by the research methodology, data analysis, hypotheses testing, and results. The paper concludes with discussing the key findings and their conclusions, suggesting future research directions.

## 2. E-Banking in Jordan

Jordan is one of the regionally leading countries regarding the national IT infrastructure available for online services. Also, the population's motivation and ability to conduct online transactions are one of the highest regionally (Alawneh and Hattab, 2008).

In Jordan, the existence of personal computers in homes has become increasingly essential to modern life

(according to a recent survey of the Ministry of Communication and Information Technology, it is estimated that 57% of Jordanian families do have a PC at home). Also, computer literacy has become mandatory in all Jordanian schools and universities (Al Nagi and Hamdan, 2009). One of the most important assets of Jordan is its population. Comparing Jordan with other countries in the region, it has the highest rates of college educated people. Also, about 50% of the population is between the ages of 15 and 30, which makes them the future generation that will become the foundation in building an information society (Intaj, 2003). Another advantage of Jordan is its small size and population, which allows changes to be spread quickly. When its size is considered in conjunction with its stable political situation, it is clear that there are definite opportunities for Jordan to quickly become a significant leader in the building of an information society within the Middle East (NIC, 2001).

Many Jordanians have Internet access at home or work, but that doesn't apply on all types of citizens and this access often depends on education and income level. In 2010, a survey was conducted in Jordan by the Ministry of Information and Communication Technology and the Department of Statistics. This survey inquired into the use of technology in Jordanian homes. Its results are discussed as a measure of Jordanian people's maturity in ICT. It was found that 57% of Jordanians own PCs and about 23% have Internet access at home. 53% of Internet users connect at home, which was ranked as the first connection place.

Traditional branch-based retail banking remains the most widespread method for conducting banking transactions in Jordan as well as any other country. However, Internet technology is rapidly changing the way personal financial services are being designed and delivered. For several years, commercial banks in Jordan have tried to introduce electronic banking (e-banking) systems to improve their operations and to reduce costs. (Alawneh and Hattab, 2008).

In this Internet age, when the customer is having access to a variety of products and services it is becoming very difficult for banks to survive. In this situation, when customer inquires are not met easily or transactions are complicated, the customer will asks for new levels services, and only chose those institutions who are making a real effort to provide a high level of quality, fast and efficient service through all the bank's touch points, call centers, ATMs, voice response systems, Internet and branches. (Alawneh and Hattab, 2009).

The financial sector in Jordan has witnessed media blitzes announcing electronic banking. Banks that have implemented e-banking are showing up of being modernized; some of those that have not are drastically trying to catch up. (Alawneh and Hattab, 2009).

The banking sector is very dynamic and liberal in Jordan. Moreover, many of the commercial banks in Jordan are offering electronic services. Table 1 and Table 2 show statistical data about the surveyed Jordanian and foreign banks which include year of establishment, number of branches, ATMs and samples of their e-Banking services.

Table 1: Statistical Data about the Surveyed Jordanian Banks

| Bank Name | year of establishment | No. of branches | No. of ATMs | Samples of E-banking services |
|---|---|---|---|---|
| Arab Bank | 1930 | 80 | 139 | Arabi Online is an easy, secure Internet banking Service. |
| Cairo-Amman Bank | 1960 | 79 | 120 | CAB On-Line is an easy to use state-of-art Internet banking service. |
| The Housing Bank | 1974 | 110 | 187 | Iskan Online |
| Jordan Islamic Bank | 1979 | 72 | 94 | The I-Banking. |

Table 2: Statistical Data about the Surveyed Foreign Banks

| Bank Name | year of establishment | No. of branches | No. of ATMs | Samples of E-banking services |
|---|---|---|---|---|
| HSBC Bank | 1949 | 6 | 14 | HSBC Online Banking. |
| Egyptian Arab Land Bank | 1951 | 14 | 14 | Aqari Online |
| BLOM Bank | 2004 | 9 | 12 | eBLOM Internet Banking Service. |
| National Bank of Kuwait | 2004 | 8 | 8 | Watani Online Banking Services. |

## 3. Relationship Quality

Relationship quality refers to the overall assessment of the strength of a relationship between two parties (Palmatier et al. 2006). Drawing upon the marketing literature, relationship quality has emerged as a paradigm that indicates the extent that the customer trusts the service provider and has confidence in the service provider's future performance because the provider's past performance has been consistently satisfactory.

Relationship quality is usually conceptualized as a composite or multidimensional construct capturing the different but related facets of a relationship (Lages et al. 2005; Palmatier et al. 2006). It is conceived in this study as a higher order construct that has three distinct yet related components: trust, satisfaction and commitment. These three components have been widely referred to in relationship quality studies as definitive components of relationship quality (Vieira et al. 2008; Hsieh & Li 2008; Roberts et al. 2003; Walter et al. 2003; Ivens 2004; Ulaga and Eggert 2006; Ivens and Pardo 2007; Kempeners 1995; Crosby et al. 1990).

After a rigorous survey of a literature relevant to relationship quality, the following table illustrates the most common dimensions of relationship quality.

Table 3: The Dimensions of Relationship Quality in the literature

| Dimensions of Relationship Quality | References in Literature |
|---|---|
| Trust, satisfaction and commitment. | (Hsieh and Li, 2008) |
| Trust, satisfaction, equity. | (Boles et al., 2000) |
| Trust, commitment, Product-related or service-related quality perception. | ((Hennig-Thurau 2000) |
| Trust, satisfaction, Coordination, Power, and Conflict. | (Naudé and Buttle 2000) |
| Trust, satisfaction. | (Shamdasani and Balakrishnan 2000) |
| Trust, satisfaction and commitment | (DeWulf et al. 2001) |
| Trust, commitment, service quality | (Hennig-Thurau et al. 2001) |
| Trust, satisfaction | (Vieira 2001) |
| Trust, satisfaction and commitment | (Hennig-Thurau et al. 2002) |
| Trust, commitment | (Hewett et al. 2002 |
| Trust, satisfaction | (Parsons 2002) |
| Trust, commitment | (Wong and Sohal 2002) |
| Trust, satisfaction and commitment, Affective conflict. | (Roberts et al. 2003) |
| Trust, satisfaction and commitment | (Walter et al. 2003) |
| Trust, satisfaction and commitment | (Ivens 2004) |
| Trust, satisfaction and commitment | (Ulaga and Eggert 2006) |
| trust, satisfaction and commitment | (Ivens and Pardo 2007) |
| Trust, satisfaction and commitment, service quality. | (Rauyruen and Miller 2007) |

## 3.1 Satisfaction

The Expectation-Confirmation Theory (ECT) was proposed by Oliver (1980) to study consumer satisfaction and re-purchase behavior. The ECT theory states that consumers firstly form an initial expectation prior to purchase, and then build perceptions about the performance of the consumed product/service after a period of initial consumption. Next, consumers will decide on their level of satisfaction based on the extent to which their expectation is confirmed through comparing the actual performance of the product/service against their initial expectation of the performance. Consequently, satisfied consumers will form re-purchasing intentions.

Satisfaction can be described as an "evaluation of the perceived discrepancy between prior expectation and the actual performance of the product" (Oliver, 1999). Satisfaction is closely related to service quality and consist of both a behavioral dimension created by experience, as well as a mental dimension, created by worked up attitudes (Oliver, 1999). On the other hand, dissatisfaction among customers using electronic services might occur because of technological failure, which results in a negative perception of the functional quality of the service. Dissatisfaction might also arise from technology design problems or service design problems. This could include systems being too slow, difficulties for the user to navigate the system or problems to figure out how to log off the service (Meuter et. al., 2000). (Anderson and Srinivasan, 2003) found that trust significantly actuate the impact of satisfaction on e-commerce loyalty.

Relationship satisfaction is an affective or emotional state toward a relationship, typically evaluated cumulatively over the history of the exchange (Palmatier et al. 2006). The satisfaction component of relationship quality is about users' evaluation of *the relationship* with the service provider. The major source of relationship satisfaction is a history of positive interaction with the service provider. The customer's best assurance of future performance is a continuous history of personalized, error-free interaction (Crosby et al. 1990).

For the purposes of this study, satisfaction dimension is defined as the customers' positive emotional state resulting from utilizing E-banking services that are provided by the Jordanian and foreign banks.

## 3.2 Trust

According to the theory of reasoned action (TRA), an individual's belief towards a behavior is an immediate determinant of his intention to perform a behavior (Fishbein and Ajzen, 1975). (Mayer et al., 1995) have extended the TRA to support the modeling of customer trust. Based on assumptions of the Technology Acceptance Model (TAM) (Davis, 1989), the intention to accept or use a new technology is determined by its perceived usefulness and perceived ease of use. (Mcknight et al., 2002) proposed a model of e-commerce customer trust where they posited that trusting beliefs leads to trusting intentions, which in turn influences trust related behaviors.

Trust is a crucial ingredient of any successful e-Commerce project success, and privacy is the key element in building citizen's trust in e- Commerce service (kim et al., 2009). By making customers interact with e- Banking as well as entrust sensitive personal and financial data to the bank, a better relationship is maintained.

Customer trust can be defined as a set of beliefs held by an online consumer concerning certain characteristics of the e-supplier, as well as the possible behavior of the e-supplier in the future ( Coulter and Coulter, 2002). Lee and Lin (2005) suggested that trust encourages online purchasing and affects customer attitudes towards purchasing from t-retailers. Kim et al. (2009) conducted a longitudinal study in the U.S and found that online customer trust is strongly related to loyalty.

Trust is an important indicator of relationship quality. Only when a person trusts the trustee will he/she be likely to perceive that there is a high quality relationship between the trustee and him/her. A relationship that lacks trust is unlikely to be perceived as of high quality. It is also important to note that trust is difficult to foster, can be shaken easily, and once shaken, is extremely difficult to rebuild (Shneiderman 2000).

For the purposes of this study, trust dimension is defined as the customers' willingness to rely on the banks that are providing E-banking services for conducting various banking activities and transactions.

## 3.3 Commitment

Defined as the enduring desire to maintain a valued relationship (Palmatier et al. 2006), customer commitment is one of the most commonly studied key determinant in relationship quality studies (Hsieh & Li 2008). Commitment has often been conceptualized as a multi-dimensional construct. For instance, Gundlach et al (1995) proposed three components of commitment: an instrument component of some form of investment, an attitudinal component described as affective commitment or psychological attachment, and a temporal dimension indicating that the relationship exists over time. Thus, a committed customer has invested time and emotion into the relationship with a service provider, shows affect toward it, and is willing to maintain this relationship for a certain period of time, sometimes even at the cost of short-term benefits.

Commitment is an essential ingredient for successful long-term relationships because committed customers are the basis for business continuity and bring future value or benefits to those they are committed to (Lemon et al. 2001). Committed customers feel loyal to the service provider and are willing to put extra efforts or even sacrifice short-term benefits to maintain the relationship. They are also more tolerant of minor errors from the service provider. Maintaining this commitment to a specific marketplace is especially important in e-commerce, given the low cost of switching to different online marketplaces.

For the purposes of this study, commitment dimension is defined as the customers' future intention to return indicating that the relationship exists over time to predict the continuity of the relationship with the banks that are providing E-banking services.

## 4. Previous Studies

Sun (2010) investigated how the quality of certain attributes of e-commerce systems — such as information quality, system quality, and service quality — can be leveraged to enhance business benefits as indicated by customer commitment and customer retention. This study argues that relationship quality, a concept encapsulating the ideas of both trust and satisfaction, is crucial for transferring attributes of e-commerce systems into business benefits. A research model of relationship quality in e-commerce was built, drawing upon information systems and marketing literature. This model was then examined using a survey of 140 online auction sellers at uBid.com. The empirical results confirmed the research model. Information quality, system quality, and service quality affect relationship quality significantly. Relationship quality in turn has significant impact on customer commitment and customer retention.

Vieira et al. (2008) reviewed a literature and research on relationship quality (RQ); their study provided a systematization of the current knowledge on RQ and offered suggestions for future research. Specifically, it reviews and synthesizes existing research on RQ and argues for a framework in which trust, satisfaction, and commitment are the three key dimensions of RQ, while mutual goals, communication, domain expertise, and relational value should be seen as core determinants which, in future models, may be augmented by context specific influences.

Chakrabarty et al. (2008) conducted a national survey of firms that participated in outsourcing relationships, and service quality and relationship quality were found to be significantly and positively related to each other and both had a significant impact on user satisfaction. However, the intricacies of the causal effects between the two autonomous constructs, service quality and relationship quality, are a source of interest. In post-analysis theory building, we give a conceptual model that proposes that the positive causal effect of service quality on relationship quality would be positively moderated by the client orientation and promotion effectiveness of the vendor, while the positive causal effect of relationship quality on service quality would be mediated by the project management effectiveness. Hence, this paper comprises of two related parts: first an empirical study, and secondly developing a theory and conceptual model that delve into the causalities involved in service quality, relationship quality, and the role of Internet technologies and collaboration tools.

Lee et al. (2011) investigated the relationships between service quality, relationship quality, and customer loyalty while deregulation of financial institutions has increased competition in the Taiwanese banking industry, the advent of e-commerce has provided business opportunities for consumer financing operations. Network banking helps banks to develop relationship marketing by delegating tasks to customers, thus improving customer loyalty. It was found that crisis handling and relationships are negatively, and relationship quality and customer loyalty, and service quality and customer loyalty positively, correlated. Customer loyalty in Taiwanese Internet banks can be increased by improving service quality and relationship quality.

Vieira (2009) study aimed to provide a better understanding of the nature and determinants of relationship quality (RQ). A combination of a literature review and the results of a series of interviews with relationship managers of hotels operating in Portugal and their counterparts in corporate clients informed the development of a model, which was subsequently, tested using LISREL. Findings highlighted the association between the presence of designated client managers and high quality business relationships, and suggested directions to improve the performance of organizations in building B2B RQ.

## 5. The Research Hypotheses

This section discusses the related hypotheses based on the literature review and the related previous works by researchers. To examine the points previously discussed and address the issues raised, we have formulated the following nine hypotheses.

H1: Customers' level of trust with e-banking services offered by Jordanian banks has a positive effect on customers' perception of their relationship quality with the bank.

H2: Customers' level of satisfaction with e-banking services offered by Jordanian banks has a positive effect on customers' perception of their relationship quality with the bank.

H3: Customers' level of commitment with e-banking services offered by Jordanian banks has a positive effect on customers' perception of their relationship quality with the bank.

H4: Customers' level of trust with e-banking services offered by foreign banks has a positive effect on customers' perception of their relationship quality with the bank.

H5: Customers' level of satisfaction with e-banking services offered by foreign banks has a positive effect on customers' perception of their relationship quality with the bank.

H6: Customers' level of commitment with e-banking services offered by foreign banks has a positive effect on customers' perception of their relationship quality with the bank.

H7: There is not a significant statistical difference regarding relationship quality in terms of trust between Jordanian and foreign banks.

H8: There is not a significant statistical difference regarding relationship quality in terms of satisfaction between Jordanian and foreign banks.

H9: There is not a significant statistical difference regarding relationship quality in terms of commitment between Jordanian and foreign banks.

## 6. The Research Methodology

Based on the literature review and previous works, A survey questionnaire was made by the researchers consisting of (24) questions about the respondents and their perception of the level of trust, satisfaction and commitment with e-Banking services offered by Jordanian and foreign banks. The survey questionnaire utilized constructs measure using multiple items, and all of the scale items represented in the survey instrument utilized a seven-point categorical rating scale. The questionnaire consisted of two parts. The first part included customers' demographic characteristics, including gender, age, type of work, educational level. A convenient sample of 350 customers from four Jordanian banks and four foreign banks at the Amman city was selected as a unit of analysis (Table 1 and Table 2). The questionnaire was administered face-to-face to the customers. The total number of returned questionnaires was 318 in a response rate of 91%. Among the collected questionnaires, 32 have missed responses resulted in 286 usable questionnaires. The second part aimed to investigate customers' opinions about dimensions of relationship quality between customers and banks.

For the purpose of data analysis, the descriptive statistics, structural equation modeling and t-test for equality of means analyses have been used and the WarpPLS Version 2.0) and SPSS 17.0 software was employed.

The internal consistency measures (Cronbach's alpha) are obtained in order to assess the reliability of the measurement constructs.

## 7. Research Findings:

### 7.1 Data Analysis

The descriptive statistics of the respondents were analyzed in Table 4, and presents the demographic characteristics of the 286 respondents. The following table gives a general overview of the sample surveyed in term of the demographic information.

The respondents' gender is distributed almost evenly between male and females. The majority of respondents' age is less than 35 years (i.e., young) and their percentage is 44.7%. The highest majority of respondents are educated, 90.6% had a scientific degree. More than a third of respondents are private sector employees and their percentage is 41.6%.

Table 4: Demographic Data for the Main Survey

| Demographic object | The valid items | No. of Respondents | Percent % |
|---|---|---|---|
| Type of Bank | Jordanian | 156 | 54.5 |
| | Foreign | 130 | 45.5 |
| Gender | Female | 139 | 48.6 |
| | Male | 147 | 51.4 |
| Educational Level | General Secondary or less | 27 | 9.4 |
| | College | 50 | 17.5 |
| | Bachelor | 95 | 33.2 |
| | Master | 65 | 22.7 |
| | Doctoral | 49 | 17.1 |
| Age | <=25 | 23 | 8.0 |
| | 26-35 | 105 | 36.7 |
| | 36-45 | 48 | 16.8 |
| | 46-55 | 55 | 19.2 |
| | >=55 | 55 | 19.2 |
| Type of Work | Public Servant | 76 | 26.6 |
| | Private Sector Employee | 119 | 41.6 |
| | Businessman | 51 | 17.8 |
| | Student | 40 | 14.0 |

The internal consistency measures for all three constructs were conducted using Cronbach alpha tests. Table 5 presents the alpha values for all factors which vary from 0.94 to 0.95 which are considered acceptable for this type of study.

Table 5: Cronbach Alpha Coefficient for Determinants of Relationship Quality

| The Dimension | No. of Items | Alpha Coefficient |
|---|---|---|
| Satisfaction | 7 | 0.940 |
| Trust | 7 | 0.954 |
| Commitment | 10 | 0.958 |
| All Dimensions of Relationship Quality | 24 | 0.968 |

7.2 Hypotheses Testing Results

The structural equation modeling (SEM) approach was used to validate the research hypotheses. Partial least squares technique (WarpPLS Version 2.0) was employed to perform the analysis. WarpPLS employs a regression resampling method for estimation. For each hypothesis, a model of regression was run separately for each of the independent variables (trust, satisfaction, and commitment). Accordingly, we examined the coefficients of the causal relationships between constructs, which would validate the hypothesized effects. Figure 1 illustrates the paths and their significance on the structural model. As shown in Figure 1, all paths are significant at the 0.01 level.

Approximately, 92% and 96% of the variance of relationship quality level in the customers of Jordanian banks group and customers of foreign banks group, respectively ($R^2$ = 0.92, 0.96) is explained by trust, satisfaction, and commitment. For Jordanian banks, the results show that all the dimensions explain 92% of the variance in relationship quality level. This means that most of the 92% of the variance in relationship quality level is attributed to trust, satisfaction, and commitment. The remaining 8% of the variance in relationship quality level might be attributed to other factors such as, quality of e-banking services, demographics of customers, online customers' behavior, banks' economic strength, cultural factors, economical factors,… etc. For foreign banks, the results show that all the dimensions explain 96% of the variance in relationship quality level. This means that most of the 96% of the variance in relationship quality level is attributed to trust, satisfaction, and commitment. The remaining 4% of the variance in relationship quality level might be attributed to other factors such as, other factors such as, quality of e-banking services, demographics of customers, online customers' behavior, banks' economic strength, cultural factors, economical factors,… etc.

The results show that trust, satisfaction and commitment determinants affect positively and significantly the relationship quality (p<0. 01), thus supporting H1, H2 and H3. As shown in Figure 1, trust dimension explains 46% of the variance in relationship quality between Jordanian banks and its customers; satisfaction dimension explain 31% of the variance in relationship quality between Jordanian banks and its customers, and commitment dimension explains 35% of the variance in relationship quality between Jordanian banks and its customers. For Jordanian banks, that means most of the 92% of the variance in relationship quality is attributed to trust, satisfaction and commitment.

Furthermore, as shown in Figure 1, the hypotheses H4, H5 and H6 were supported in the testing. Specifically, trust dimension explains 40% of the variance in relationship quality between foreign banks and its customers; satisfaction dimension explain 32% of the variance in relationship quality between foreign banks and its customers, and commitment dimension explains 34% of the variance in relationship quality between foreign banks and its customers. For foreign banks, that means most of the 96% of the variance in relationship quality is attributed to trust, satisfaction and commitment.

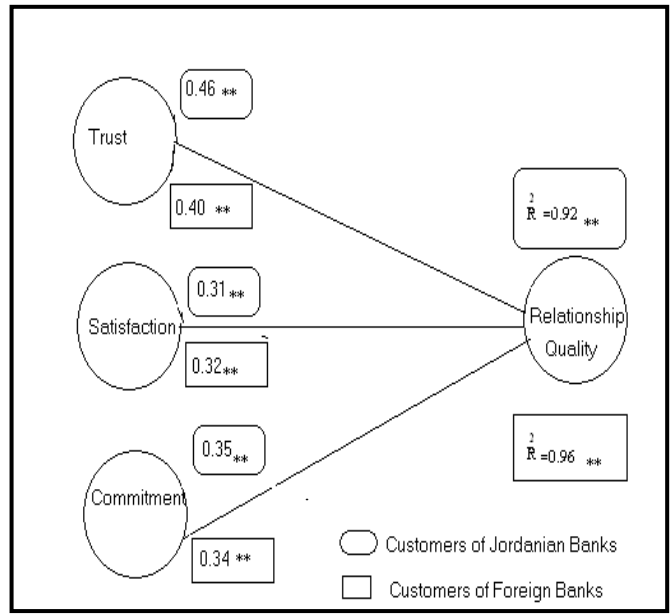

Fig. 1 Path diagram for research hypotheses

Additionally, we analysed customers' perceptions upon the three dimensions. Table 6 shows the mean scores and standard deviations between the two groups. Table 7 shows the mean difference and t-value, together with significant 2-tailed ratios between the two groups, i.e.

customers of Jordanian banks and customers of foreign banks. These differences in perceptions of customers upon the three dimensions in the two groups were examined using t-test for Equality of Means. As shown in above Tables, significant differences were found for trust, satisfaction and commitment based on the bank (i.e. Jordanian or foreign). These results indicate that customers of Jordanian banks have a higher positive perception about the trust, satisfaction and commitment than customers of foreign banks. Thus, H7, H8 and H9 were rejected.

Table 6 Group Statistics

| | bank | N | Mean | Std. Deviation | Std. Error Mean |
|---|---|---|---|---|---|
| satisfaction | Jordanian | 156 | 5.5797 | 1.00273 | .08028 |
| | Foreign | 130 | 5.1670 | 1.00939 | .08853 |
| trust | Jordanian | 156 | 5.4487 | 1.14980 | .09206 |
| | Foreign | 130 | 4.9934 | 1.16383 | .10207 |
| commitment | Jordanian | 156 | 5.0679 | .95802 | .07670 |
| | Foreign | 130 | 4.7331 | .92396 | .08104 |
| RQ | Jordanian | 156 | 5.3654 | .90302 | .07230 |
| | Foreign | 130 | 4.9645 | .91331 | .08010 |

Table 7 t-test for Equality of Means

| Dimension | Mean Difference | Std. Error Difference | df | T | Sig. (2-tailed) |
|---|---|---|---|---|---|
| Satisfaction | .41264 | .11944 | 284 | 3.455 | .001 |
| Trust | .45531 | .13730 | 284 | 3.316 | .001 |
| Commitment | .33487 | .11195 | 284 | 2.991 | .003 |
| Relationship quality | .40094 | .10779 | 284 | 3.720 | .000 |

## 8. Discussion of Findings

The results showed that trust, satisfaction, and commitment had a significant influence on relationship quality level for those customers who are transacting with both Jordanian and foreign banks that are providing E-banking services for conducting various banking activities and transactions.

The research findings showed that the hypotheses 1, 2, 3, 4, 5 and 6 are fully proved through the significance of the three determinants: trust, which is related to the customers' willingness to rely on the Jordanian or foreign banks that are providing E-banking services for conducting various banking activities and transactions; satisfaction which is related to the customers' positive emotional state resulting from utilizing E-banking services that are provided by the Jordanian or foreign banks; commitment which is related to the customers' future intention to return indicating that the relationship exists over time to predict the continuity of the relationship with the Jordanian or foreign banks that are providing E-banking services.

Accordingly, trust factor contributes significantly and positively in enhancing the relationship quality between Jordanian banks and its customers; it accounts for 46% of the change and variance in relationship quality. Furthermore, it contributes significantly and positively in enhancing the relationship quality between foreign banks and its customers; it accounts for 40% of the change and variance in relationship quality. This indicates the significant role of the trust in influencing customers' perception of their relationship with the two groups of banks. This means that the higher ability to rely on the bank's behavior, the more expectations of the customer will be met. In this sense, the trust will entice customers to fully transform to e-banking services and perform their banking and financial transactions online. These findings are consistent with the findings of Sun (2010), Vieira et al., (2008), Vieira (2009), Lee et al., (2011), Chakrabarty et al., (2008), and Skarmeas & Robson, (2008).

Satisfaction factor contributes significantly and positively in enhancing the relationship quality between Jordanian banks and its customers; it accounts for 31% of the change and variance in relationship quality. Furthermore, it contributes significantly and positively in enhancing the relationship quality between foreign banks and its customers; it accounts for 32% of the change and variance in relationship quality. This indicates the significant role of the satisfaction in achieving assurance perceived by customers regarding the bank's future performance because the level of past performance has been consistently satisfactory with the two groups of banks. These findings are consistent with the findings of Sun (2010), Vieira et al., (2008), Vieira (2009), Lee et al., (2011), Chakrabarty et al., (2008), Skarmeas & Robson, (2008).

Commitment factor contributes significantly and positively in enhancing the relationship quality between Jordanian banks and its customers; it accounts for 35% of the change and variance in relationship quality. Furthermore, it contributes significantly and positively in enhancing the relationship quality between foreign banks and its customers; it accounts for 34% of the change and variance in relationship quality. This indicates the significant role of the commitment in creation of a consistent motivation between the two groups of banks and their customers to maintain a relationship, which is seen as essential for each party for achieving its goals. These findings are consistent with the findings of Sun (2010), Vieira et al., (2008), Vieira (2009), Lee et al., (2011), Chakrabarty et al., (2008), Skarmeas & Robson, (2008).

In that respect, as shown in Figure 1; the trust factor is the strongest determinant of relationship quality level for customers of Jordanian banks (0.46) than other determinants satisfaction, and commitment (0.31, 0.35) respectively. Also, the trust factor is the strongest determinant of relationship quality level for customers of foreign banks (0.40) than other determinants satisfaction, and commitment (0.32, 0.34) respectively. As regards our study, the three determinants satisfaction, commitment, and trust have found to be critical factors for relationship quality level with Jordanian and foreign banks that are providing e-banking services.

Additionally, the results of the hypotheses testing verified that there are significant statistical difference regarding relationship quality in terms of trust, satisfaction and commitment between Jordanian and foreign banks. As shown in Table 4 and Table 5, these results indicate that customers of Jordanian banks have a higher positive perception about the trust, satisfaction and commitment than customers of foreign banks. Possible explanations for these findings are as follows: First, due to the fact that the majority of customers are Jordanians, hence they trust highly Jordanian banks more than foreign banks. This might be attributed to their previous experiences with the bank, demographics of Jordanians and the financial and economical stability of Jordanian banks. Second, considering that Jordanian banks are more customized, localized and oriented toward achieving the needs, desires and interests of Jordanians, hence Jordanians are more committed with Jordanian banks than foreign. This is probably due to previous experiences with the Jordanian banks in terms of providing distinctive services and information, as well as building long-term relationships with Jordanians. Third, given that foreign banks have huge financial capital, international banking expertise and advanced ICT infrastructure which in turn led to providing high quality services. This has been resulted in making Jordanians more satisfied with services of foreign banks than Jordanian banks.

## 9. Conclusions and Future Works

This study presents important implications for research and practice. In spite of the rapid growth of e-Banking, previous research has mostly been limited to the issues of adoption, acceptance and usage of e-Banking services. This study concentrated instead on relationship quality between customers and banks that are providing e-Banking services.

The findings of this study have important implications for banking practitioners in Jordanian and foreign banks by providing strategic insights into achieving high levels of relationship quality between banks and their customers who are utilizing the e-Banking services.

First, with the help of this study the authors managed to determine those determinants whereupon the relationship quality is based and qualitative mutual relations that develop between the bank (i.e., Jordanian or foreign) and its customers. results obtained in the study also acknowledged how the bank (i.e., Jordanian or foreign )can create excellent quality of relations: by meeting the expectations of its clients, maintaining on its promises to its customers, taking into account the interests of its clients, providing information about its services continuously, clarifying and supporting the right decision for its clients, fulfilling its commitments to its customers, continuing to provide high quality services, producing excellent service quality, providing up-to-date information for the bank's services and products, providing competent information and solving conflict situations in a professional way.

Second, this study reveals the current quality of relationship in terms of customers' satisfaction, trust and commitment that may be used in customer relationship management that allows obtaining information on how to attract a customer in a possibly more efficient way and develop good mutual relationship with it. Of course, the obtained study will be of great expedience to the bank, wherein the study was carried out for better understanding of the current customers' satisfaction, trust and commitment and activities to be performed to improve present showings and make them perfect.

Finally, both the banks (i.e., Jordanian or foreign) and customers must make efforts to improve the relationship quality through various means like increased formal socialization (meetings and conferences) and informal socialization (parties, get-togethers, and joint celebrations of success), and therefore attempt to improve the atmosphere of trust, commitment, satisfaction, cultural tolerance, interdependencies and communication.

Accordingly, high relationship quality would translate to high service quality only through effective communication between bank (i.e., Jordanian or foreign) and its customers. On the other hand, poor relationship quality (displayed through frequent conflicts and non-cooperation) would lead to ineffective communication that would ultimately hamper service quality of the bank.

This study is not without limitations, the sample size is not large enough. It is only from four Jordanian banks and four foreign banks working in Amman the capital of Jordan. Therefore, to increase generalization and accuracy of the study findings future studies should attempt a larger sample size to include participants from all regions and all banks' branches in Jordan.

The current research is limited to one application E-banking. Nonetheless, other E-applications in Jordan such as E-government, E-learning, E-tailing …etc can be studied and so this would improve the generalizability of the research findings.

The findings of this study, however, should be interpreted with the following limitations in mind. Since the proposed hypotheses of this study are tested on Jordanian customers, the generalizability of the findings could be limited to the Jordanian social- economical-cultural contexts. Therefore, cross-cultural or sub-cultural comparative studies, as well as replication research in different services contexts, should be conducted in the future to provide a useful empirical basis to enhance the external validity of the findings.

**Dr. Ali Alawneh** is an assistant professor and a head of Management Information Systems department at Philadelphia University, Jordan. He got his PhD from Arab Academy for Banking and Financial Sciences, 2008. His PhD involved a study in e-business in which he investigated e-business value creation at Jordanian Banking Sector through developing a model named e-TOEECLN. His research interests include technology diffusion, e-business models and management, e-commerce, e-banking, information systems development, e-learning and knowledge management. Dr. Alawneh has 15 publications in Scientific Conferences and Journals in the area of Information Technology, e-business, e-banking and Knowledge Management. He is a member of the technical committee of The International Arab Journal of e-Technology (IAJET), Excellent constructive reviewer- International Business Information Management Conference (12 IBIMA) - Malaysia. Member of technical committee (Reviewer)- ISIICT 2009 Third International Symposium on Innovation in Information & Communication Technology 15 - 17 December, 2009, Philadelphia University, Amman, Jordan http://www.philadelphia.edu.jo/isiict2009. Excellent constructive reviewer- International Business Information Management Conference (13 IBIMA) - Morocco. Member of Organizing committee in ISIICT 2009 Third International Symposium on Innovation in Information & Communication Technology 15 - 17 December, 2009, Philadelphia University, Amman, Jordan http://www.philadelphia.edu.jo/isiict2009. Member of the steering committee of the fourth International Symposium on Innovation in Information & Communication Technology - ISIICT 2011 November, 2011